\documentclass[
]{ceurart}

\usepackage{listings}
\usepackage{booktabs}
\usepackage{graphicx}
\usepackage{subcaption}
\usepackage{float}

\begin{document}

\copyrightyear{2025}
\copyrightclause{Copyright for this paper by its authors.
  Use permitted under Creative Commons License Attribution 4.0
  International (CC BY 4.0).}

\conference{}

\title{The ISLab Solution to the Algonauts Challenge 2025: A Multimodal Deep Learning Approach to Brain Response Prediction}


\author{Andrea Corsico}[email=a.corsico@campus.unimib.it]\cormark[1]
\author{Giorgia Rigamonti}[orcid=0009-0006-4253-1020,email=giorgia.rigamonti@unimib.it]
\author{Simone Zini}[orcid=0000-0002-8505-1581,email=simone.zini@unimib.it]
\author{Luigi Celona}[orcid=0000-0002-5925-2646,email=luigi.celona@unimib.it]
\author{Paolo Napoletano}[orcid=0000-0001-9112-0574,email=paolo.napoletano@unimib.it]

\cortext[1]{Corresponding author.}
\address{Department of Informatics, Systems and Communication, University of Milano-Bicocca, viale Sarca 336 -- 20126, Milano, Italy}




\begin{abstract}
In this work, we present a network-specific approach for predicting brain responses to complex multimodal movies, leveraging the Yeo 7-network parcellation of the Schaefer atlas. Rather than treating the brain as a homogeneous system, we grouped the seven functional networks into four clusters and trained separate multi-subject, multi-layer perceptron (MLP) models for each. This architecture supports cluster-specific optimization and adaptive memory modeling, allowing each model to adjust temporal dynamics and modality weighting based on the functional role of its target network. Our results demonstrate that this clustered strategy significantly enhances prediction accuracy across the 1,000 cortical regions of the Schaefer atlas. The final model achieved an eighth-place ranking in the Algonauts Project 2025 Challenge, with out-of-distribution (OOD) correlation scores nearly double those of the baseline model used in the selection phase. Code is available at https://github.com/Corsi01/algo2025.
\end{abstract}


\maketitle

\section{Introduction}
%
%
%
%
A central goal of computational neuroscience is to model how the brain responds to naturalistic stimuli. Traditional brain encoding models have focused on individual sensory modalities using controlled laboratory stimuli, achieving success in predicting neural responses within specific cortical regions. However, real-world perception involves the simultaneous integration of visual, auditory, and linguistic information across distributed brain networks.
Recent advances in deep learning have provided powerful tools for brain modeling. Early linear encoding models demonstrated that neural responses could be predicted from features extracted by neural networks. Since then, more sophisticated methods have emerged, including transformer architectures and multimodal models capable of processing diverse sensory inputs in parallel.
The development of large-scale neuroimaging datasets has enabled brain modeling. While earlier datasets were often limited in scope or modality, the CNeuroMod dataset provides extensive fMRI recordings of brain responses to naturalistic movie stimuli, offering an unprecedented opportunity for building and evaluating whole-brain encoding models.
The Algonauts Project 2025 Challenge \cite{gifford2025algonautsproject2025challenge} leverages this dataset to assess computational models based on their ability to predict brain responses to multimodal movie content and generalize across stimulus distributions.

\section{Background and Related Work}
\begin{figure}[htbp]
    \centering
    \begin{subfigure}{0.43\textwidth}
        \centering
        \includegraphics[width=\textwidth]{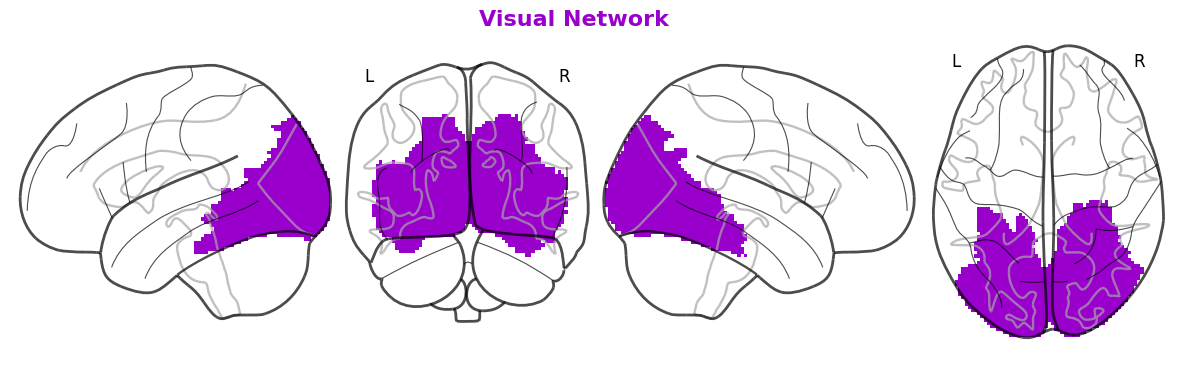}
        \label{fig:sub1}
    \end{subfigure}
    \hfill
    \begin{subfigure}{0.43\textwidth}
        \centering
        \includegraphics[width=\textwidth]{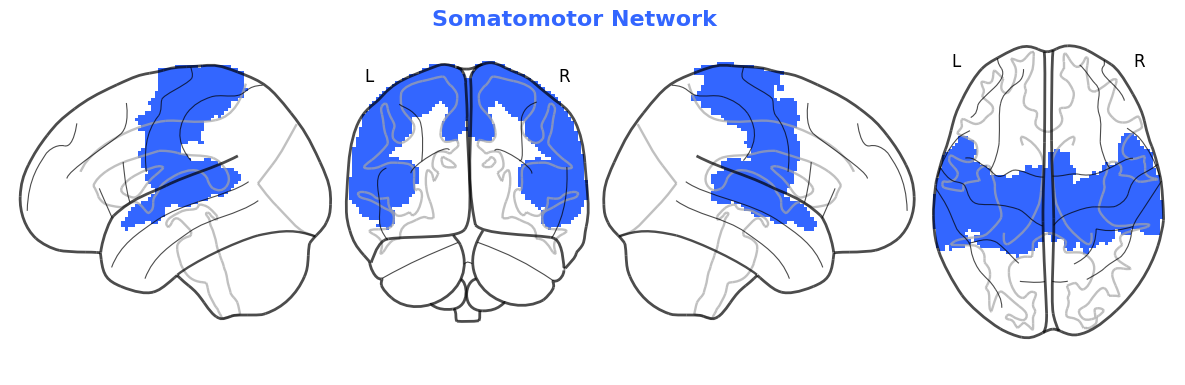}
        \label{fig:sub2}
    \end{subfigure}
    
    \begin{subfigure}{0.43\textwidth}
        \centering
        \includegraphics[width=\textwidth]{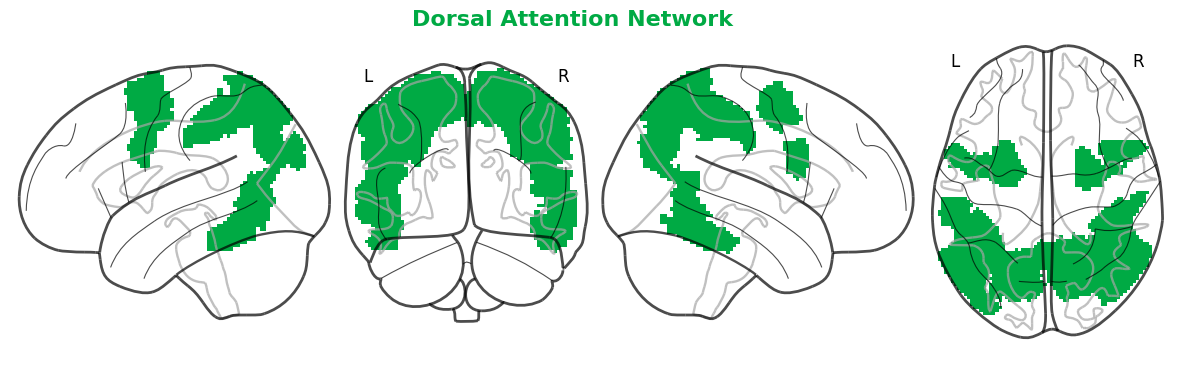}
        \label{fig:sub3}
    \end{subfigure}
    \hfill
    \begin{subfigure}{0.43\textwidth}
        \centering
        \includegraphics[width=\textwidth]{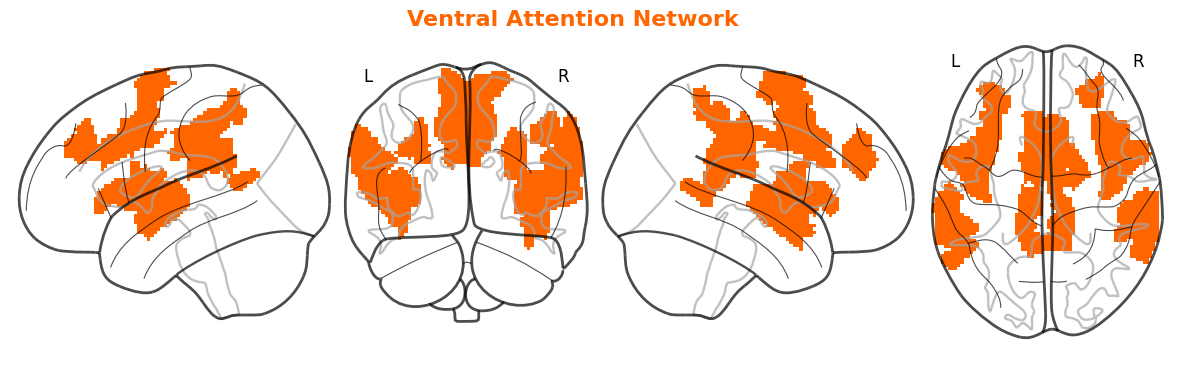}
        \label{fig:sub4}
    \end{subfigure}
    
    \begin{subfigure}{0.43\textwidth}
        \centering
        \includegraphics[width=\textwidth]{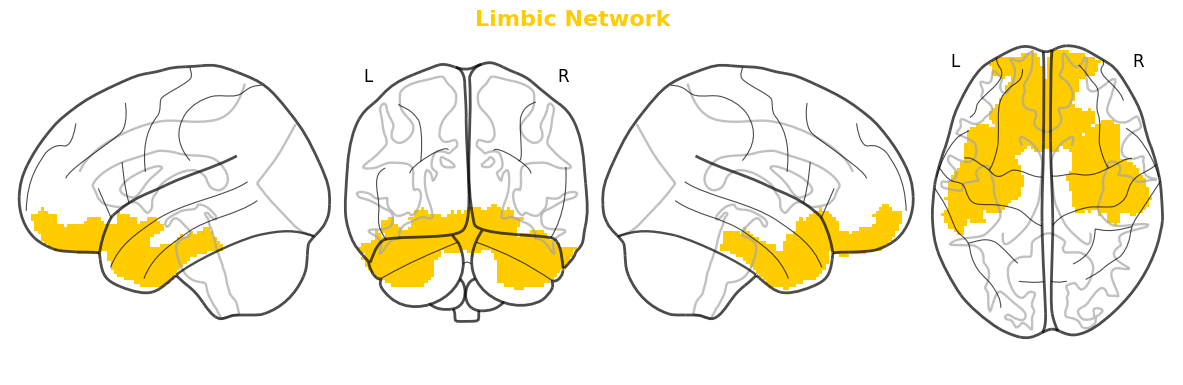}
        \label{fig:sub5}
    \end{subfigure}
    \hfill
    \begin{subfigure}{0.43\textwidth}
        \centering
        \includegraphics[width=\textwidth]{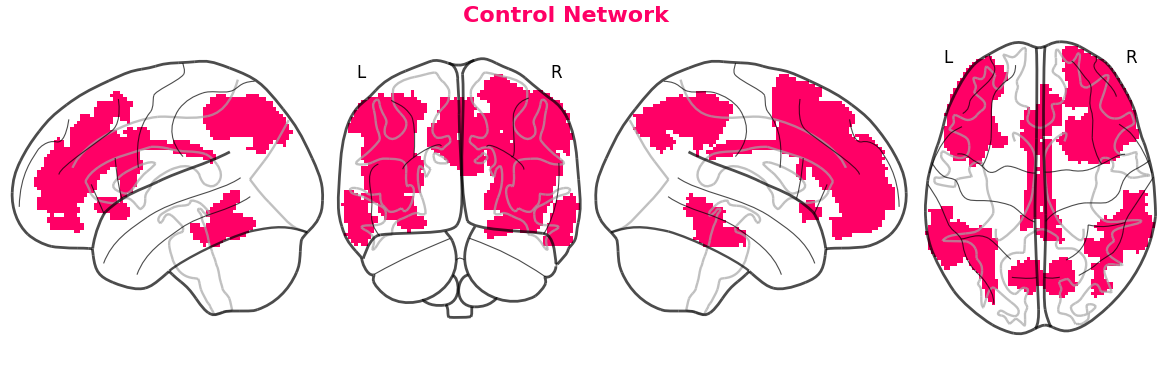}
        \label{fig:sub6}
    \end{subfigure}
    \begin{subfigure}{0.43\textwidth}
        \centering
        \includegraphics[width=\textwidth]{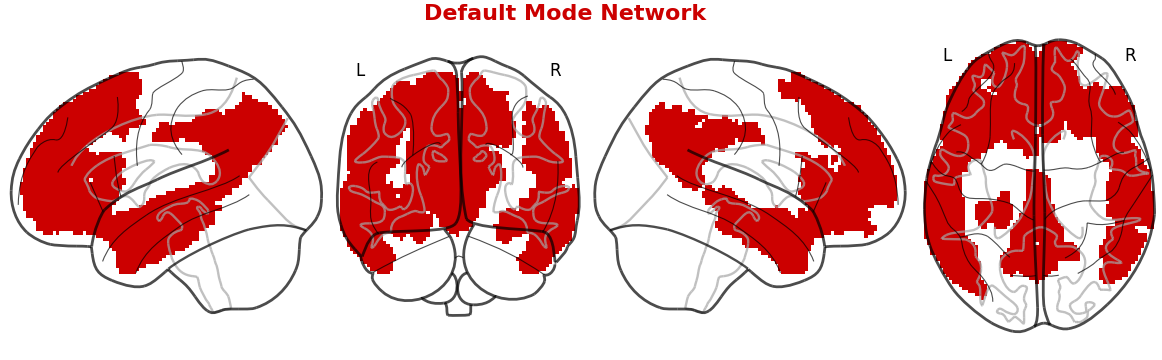}
        \label{fig:sub7}
    \end{subfigure}
    \caption{\textit{Yeo 7-network parcellation of Schaefer atlas.}}
    \label{fig:encoding_accuracy}
\end{figure}

Recent work in the Algonauts Project has yielded two key insights that, while developed for earlier challenge formats with more limited spatial coverage, offer valuable guidance for modeling whole-brain responses to multimodal stimuli.
First, Yang et al. \cite{yang2023memoryencodingmodel} demonstrated the importance of modeling temporal dynamics and incorporating memory components into brain encoding models. Their results showed that using information from past stimuli—rather than relying solely on the current time point—substantially improves prediction accuracy, highlighting the role of temporal context in neural processing.
Second, Nguyen et al. \cite{nguyen2023algonautsproject2023challenge} showed that training models across multiple subjects, followed by subject-specific fine-tuning, provides a powerful way to leverage shared patterns of neural organization while accounting for individual variability. This strategy improves generalization and robustness across subjects.
The fMRI data used in this challenge are parcellated using the Schaefer 1000-region atlas \cite{10.1093/cercor/bhx179}, which divides the cortex into 1,000 functionally defined regions, providing whole-brain coverage at spatial resolution. These regions are further organized using the Yeo 7-network parcellation \cite{yeo2011organization}, which groups them into seven large-scale functional networks. This hierarchical structure supports modeling approaches that are sensitive to both fine-grained and network-level organization (Figure \ref{fig:encoding_accuracy}).

\section{The Algonauts Challenge}
\subsection{Problem Definition}
The Algonauts Project 2025 Challenge presents a more complex brain encoding task than previous editions, introducing multimodal movie stimuli that combine visual, auditory, and linguistic information \cite{gifford2023algonauts, cichy2021algonauts}. This multimodal nature fundamentally alters the encoding problem, requiring models to integrate information across sensory modalities while predicting neural activity throughout the Schaefer 1000-region cortical parcellation.
At this scale, traditional layer-wise feature selection approaches—where model layers are matched to specific hierarchical brain regions—become computationally infeasible. Moreover, the challenge spans functionally diverse brain systems, including sensory, motor, language, attention, and default mode networks. These systems differ in their temporal dynamics, modality preferences, and information processing roles, demanding modeling frameworks that are both flexible and functionally adaptive.
A central component of the 2025 challenge is the requirement for out-of-distribution (OOD) generalization. Models are trained on episodes from the TV series Friends, but evaluated on entirely different movie content, featuring distinct visual aesthetics, narrative structures, and acoustic environments. This setting ensures that successful models learn robust principles of brain organization rather than overfitting to specific stimulus features.
Temporal processing adds further complexity. Naturalistic movie stimuli unfold over time, and brain networks exhibit varying temporal receptive windows. Primary sensory regions respond to fast-changing inputs, while higher-order association areas integrate information over longer timescales. Effective models must therefore incorporate adaptive temporal mechanisms that align with the dynamics of different cortical regions.
Finally, the challenge provides fMRI data from four subjects who viewed the same stimuli. While this limits intersubject data diversity, it emphasizes the need for models that generalize across individuals by capturing both shared neural coding principles and subject-specific patterns of brain activity.

\subsection{Data}

This challenge is based on the CNeuroMod dataset \cite{boyle2023courtois}, one of the most extensive publicly available collections of human brain responses to naturalistic movie stimuli. It provides over 80 hours of fMRI recordings from four adult participants, each exposed to identical stimulation protocols. Its scale and structure support the development of whole-brain, multimodal encoding models grounded in real-world perception.
The training set includes two primary sources: the complete Friends series (seasons 1–6) and the Movie10 dataset, which features four films—Life, The Bourne Identity, The Wolf of Wall Street, and Hidden Figures. fMRI data are provided in preprocessed form at a temporal resolution of TR = 1.49 seconds. All recordings are aligned to the Schaefer 1000-region atlas, a high-resolution cortical parcellation that divides the brain into 1,000 functionally defined regions. This atlas ensures full-brain coverage while preserving spatial granularity for capturing local neural variations.
The evaluation protocol includes two phases to assess both in-distribution (ID) and out-of-distribution (OOD) generalization. In the first phase, models are trained and evaluated on Friends season 7, preserving stimulus consistency while testing generalization across time.
The second phase tests models on a fully held-out OOD set of six diverse films: Chaplin, Princess Mononoke, Planet Earth, Passepartout, World of Tomorrow, and Pulp Fiction. These films span a wide range of content, visual styles, and language types—including English, French, and non-verbal segments; realistic and animated visuals; and settings from indoor dialogue to nature and abstract scenes.
This diversity presents a strong generalization challenge: models must go beyond surface-level stimulus features and capture broader principles of brain organization and cross-modal integration. The use of four subjects with identical stimuli also emphasizes the need to model both shared neural patterns and individual variability.

\section{Proposed Model}
\subsection{Feature extraction}
A consistent feature extraction strategy was applied across all modalities. Features were computed from 1.49-second stimulus windows, matching the fMRI repetition time (TR), to ensure precise temporal alignment between stimulus representations and corresponding neural responses. The initial extraction process produced high-dimensional embeddings for video and audio, and variable-length token sequences for text, introducing challenges related to dimensionality and cross-modal consistency. To address these issues, statistical pooling operations—including mean, maximum, and standard deviation—were applied to each modality’s features, yielding fixed-size vectors for every temporal window. These pooled representations were then reduced in dimensionality using Principal Component Analysis (PCA), facilitating efficient integration into subsequent encoding models.

\subsubsection{Visual Features}
Two approaches were employed for visual feature extraction, each targeting different aspects of perceptual relevance and spatiotemporal representation.
First, we used the ViNET model to generate saliency maps from video frames \cite{jain2021vinet}. These maps were applied to mask the video input, preserving only the most salient regions likely to capture human visual attention. The masked videos were then passed through the same model, and feature vectors were extracted from the backbone network. This strategy aimed to enhance the biological plausibility of the features by emphasizing perceptually relevant content.
Second, we utilized VideoMAE2, a transformer-based architecture designed for video understanding through masked autoencoder pretraining \cite{wang2023videomae}. Features were extracted from the final layer of the model, capturing high-level temporal and spatial dynamics across video sequences.

\subsubsection{Audio features}
To capture the diverse characteristics of naturalistic audio, three complementary approaches were employed for feature extraction. First, Wav2Vec2.0 was used to extract speech-related features, leveraging its self-supervised training on large-scale speech corpora to produce linguistically meaningful representations \cite{baevski2020wav2vec}. Second, openSMILE was applied to extract low-level acoustic features, including spectral, prosodic, and temporal properties of the audio signal \cite{eyben2010opensmile}. Third, AudioPANNs (Pre-trained Audio Neural Networks) were used to capture features associated with non-speech content such as environmental sounds and music \cite{kong2020panns}. Together, these methods provided a comprehensive representation of the auditory landscape in multimodal movie stimuli.

\subsubsection{Language Features}
Language features were extracted using RoBERTa-base, a transformer-based model pre-trained on large-scale text corpora \cite{liu2019robertarobustlyoptimizedbert}. Contextualized word embeddings were obtained from the 8th hidden layer, which has been shown to best predict brain responses in prior studies \cite{lamarre2022attention}. To generate fixed-length sentence-level representations aligned with the fMRI temporal resolution, statistical pooling operations (e.g., mean, max, and standard deviation) were applied across the word embeddings within each 1.49-second window. In addition to hidden states, attention weights from all transformer layers were also extracted and included as features. This choice was motivated by findings from Lamarre et al. (2023), who demonstrated that attention weights reliably predict language-evoked brain activity and capture aspects of contextual integration not fully represented in hidden states \cite{lamarre2022attention}. These attention patterns provide complementary information, reflecting the internal mechanisms by which the model dynamically integrates information across words.

\subsection{Multi subjects model}
To leverage data from all four subjects while accounting for differences in brain organization, we implemented a multi-subject MLP architecture. The model comprises a shared backbone network that learns common feature representations across subjects, along with subject-specific prediction heads that model neural response patterns.
The architecture incorporates trainable subject embeddings that transform subject identity from one-hot encoding to dense representations. These embeddings are concatenated with the input features and processed through a shared backbone, which extracts subject-agnostic representations that capture shared patterns of neural encoding.
Differences are modeled via separate linear heads for each subject, which map from the shared backbone representations to predicted brain responses. This design enables the model to learn both generalizable encoding mechanisms and subject-specific response characteristics, effectively increasing the diversity of the training data while maintaining the ability to capture variations in brain organization.

\subsection{Network memory modeling}
\begin{figure}[h]
    \centering
    \includegraphics[width=0.8\textwidth]{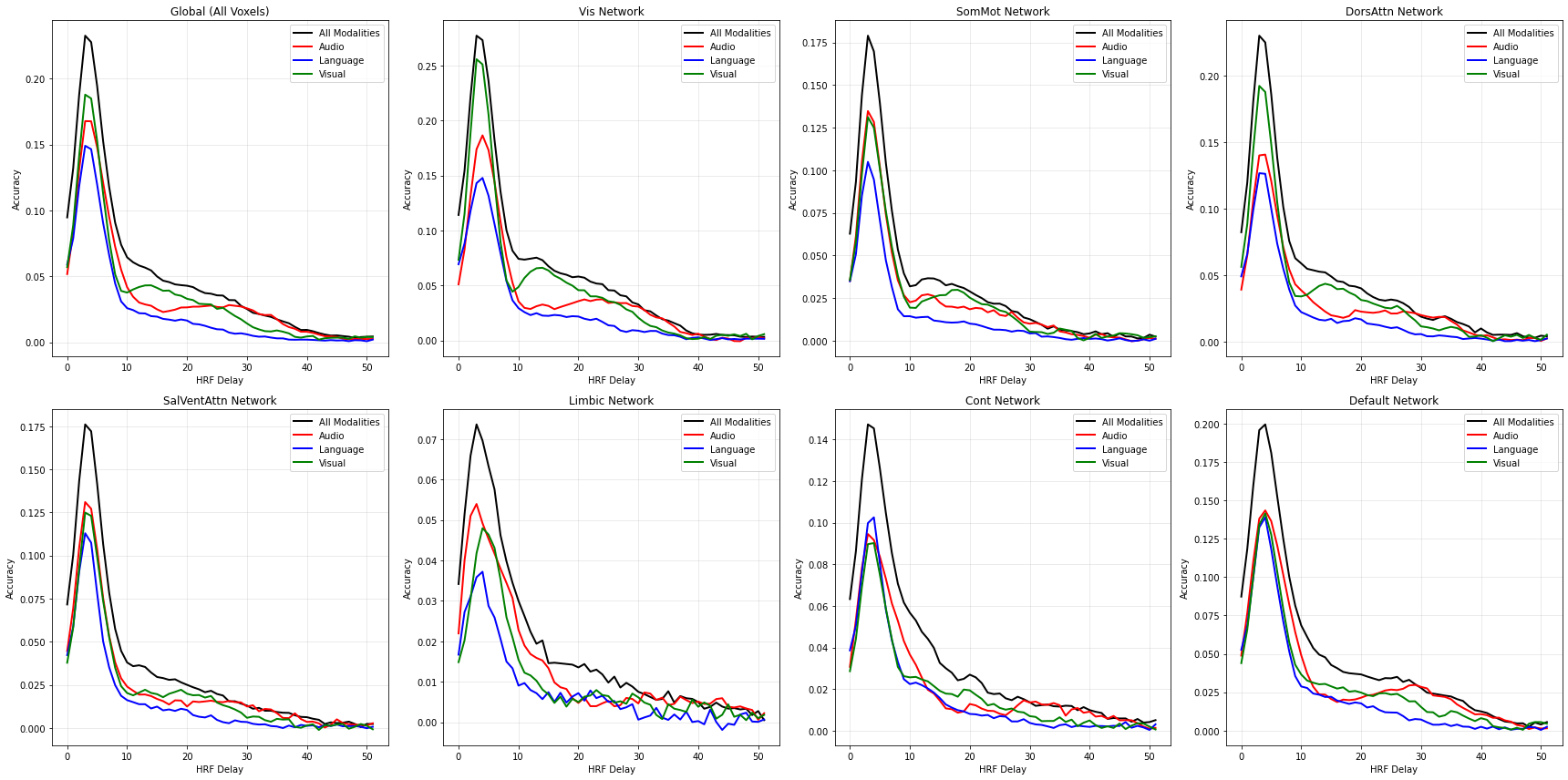}
    \caption{\textit{Temporal response patterns across Yeo networks and modalities.} Correlation performance as a function of HRF delay (0-50 time points) for individual modalities and combined features across the seven functional networks. }
    \label{fig:network_memory}
\end{figure}

To investigate the temporal dynamics of neural responses across functional brain networks, we performed a systematic analysis of memory effects by fitting models with lag windows (Figure \ref{fig:network_memory}). Each modality was tested independently across all seven Yeo networks, with results averaged across the four subjects. This analysis revealed distinct temporal response profiles for different modality-network combinations.
Based on these network-specific temporal characteristics, we adopted a data-driven strategy to incorporate additional memory features by concatenating them to the input feature vector. The exploration identified three networks that benefited significantly from memory components: Visual and Dorsal Attention networks showed improved performance with visual memory features, while the Somatomotor network benefited from both visual and audio memory features.
In response to these findings, we created four separate multi-subject MLP models: dedicated models for Visual, Somatomotor, and Dorsal Attention networks—each incorporating the appropriate memory features—and a combined model for the remaining four networks (Ventral Attention, Limbic, Frontoparietal, and Default Mode), which did not show notable memory-driven improvements. This architecture allows each model to optimize according to the temporal dynamics and modality preferences of its respective brain networks.

\section{Experiments}
The model predicts neural responses at time points independently, using a window of past stimuli to account for the delay introduced by the hemodynamic response function (HRF), which links neural activity to the BOLD signal measured by fMRI. Through systematic grid search over combinations of window lengths and delays, we selected an HRF delay of 2 time points and a temporal window of 7 time points for all modalities.
Training was performed using the Adam optimizer and a subject-weighted mean squared error (MSE) loss to address imbalances in subject representation across batches. Hyperparameters were optimized separately for the four network-specific models using Optuna \cite{akiba2019optuna}. Tuned parameters included the dimensions of the subject embedding and shared hidden layers, as well as regularization parameters such as dropout and weight decay.
Given that the two visual feature types—ViNET saliency-masked and VideoMAE2—did not perform well when used together, we trained separate models for each visual feature representation. The MLP architecture proved superior to ridge regression for brain regions but required strong regularization to prevent overfitting. All hyperparameter optimization and training procedures were conducted using Friends season 6 as the validation set. Final models were retrained on the full dataset using the optimized parameters before submission.

\subsection{Brain responses to in-distribution movies}
The model was evaluated on Friends season 7 as the ID test set. Among the two visual feature approaches, ViNET saliency-masked features consistently outperformed VideoMAE2 features across all network models, leading to the selection of ViNET-based models for final submission.
Validation on Friends season 6 revealed distinct performance patterns across the Yeo networks, supporting the effectiveness of the network-clustered modeling approach.
Correlation analyses demonstrated that the predictability of brain responses varied substantially across networks. Some networks—particularly those incorporating memory-augmented features—showed improvements in performance, while others performed comparably well using the standard (non-memory-augmented) models (Table~\ref{tab:model_performance}).
\begin{table}[ht]
\centering
\footnotesize
\setlength{\tabcolsep}{3pt}
\caption{\textit{Model performance across brain networks (validation set).} (1) Memory models, (2) No-memory models, (a) ViNet, (b) VideoMAE2. Network sizes: Visual (Vis) - 162, Somatomotor (Som) - 194, Dorsal Attention (Dors) - 122, Ventral Attention (Vent) - 121, Limbic (Limb) - 60, Default Mode (Def) - 212, Frontoparietal Control (Ctrl) - 129, Whole brain (Mean) - 1000.}
\label{tab:model_performance}
\begin{tabular}{@{}lcccccccc@{}}
\toprule
\textbf{Model} & Vis & Som & Dors & Vent & Limb & Def & Ctrl & Mean \\
\midrule
MLP (1a) & \textbf{0.390} & \textbf{0.229} & \textbf{0.310} & \textbf{0.220} & \textbf{0.113} & \textbf{0.275} & \textbf{0.231} & \textbf{0.267} \\
MLP (1b) & 0.387 & 0.223 & 0.297 & 0.214 & 0.113 & 0.274 & 0.230 & 0.263 \\
MLP (2a) & 0.363 & 0.218 & 0.301 & 0.217 & 0.109 & 0.271 & 0.229 & 0.259 \\
Ridge (2a) & 0.361 & 0.210 & 0.292 & 0.209 & 0.102 & 0.263 & 0.218 & 0.250 \\
\bottomrule
\end{tabular}
\end{table}


\subsection{Brain responses to out-of-distribution movies}
The OOD films exhibited diverse visual styles and content characteristics. Chaplin, as a silent film, contained no speech. To address this, we represented the language modality with a constant “no speech” feature vector throughout the film’s duration.
Due to the visual properties of the OOD stimuli, we adapted our visual feature extraction strategy. Since the ViNET backbone was trained on Kinetics-400 (primarily featuring common human actions), we employed VideoMAE2 features for Planet Earth (natural scenes) and Chaplin (black-and-white cinematography). This adjustment led to improved performance during the OOD evaluation phase, where VideoMAE2 outperformed ViNET for these film types.

\section{Results}
Our approach achieved competitive performance in both the in-distribution and out-of-distribution evaluation phases. 
\begin{figure}[!ht]
    \centering
    \begin{subfigure}{\textwidth}
        \centering
        \includegraphics[width=.6\textwidth]{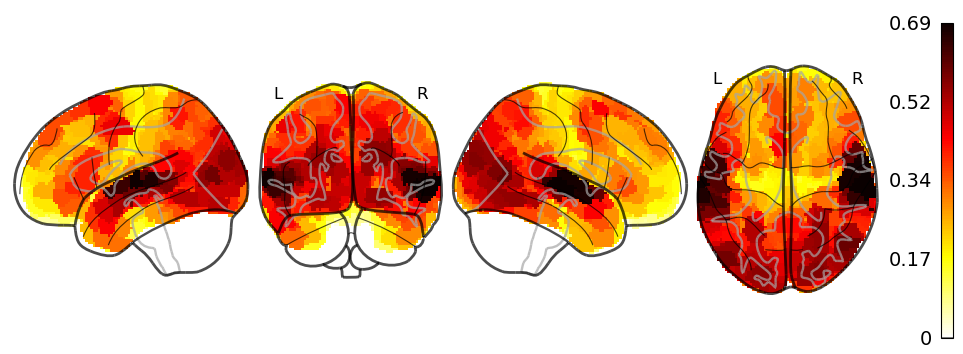}
        \caption{In-distribution: Encoding accuracy Friends s7, sub-average, mean accuracy: 0.2659}
        \label{fig:sub11}
    \end{subfigure}
    \begin{subfigure}{\textwidth}
        \centering
        \includegraphics[width=.6\textwidth]{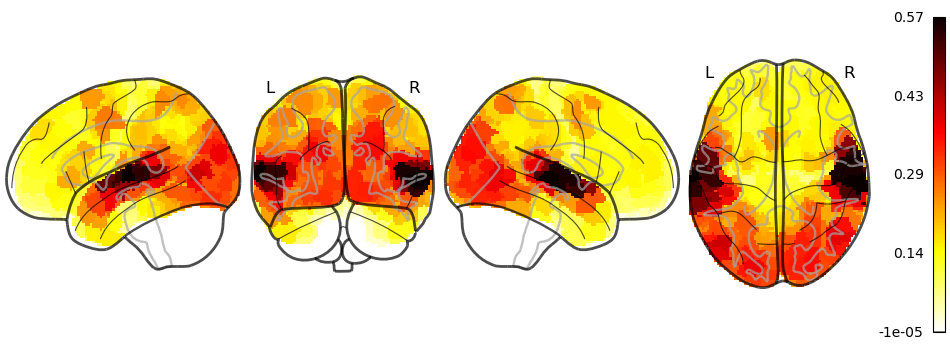}
        \caption{Out-of-distribution: Encoding accuracy OOD, subject-average, movie-average, mean accuracy: 0.1576}
        \label{fig:sub12}
    \end{subfigure}
    \caption{ \textit{Brain encoding performance comparison between in-distribution and out-of-distribution evaluation.} Correlation maps showing prediction accuracy across cortical regions, averaged across subjects (ID) and across cortical regions, movies and subjects (OOD).}
    \label{fig:combined}
\end{figure}
The model significantly outperformed the baseline, validating the effectiveness of the network-clustered, multi-subject approach for multimodal brain encoding.
Analysis of prediction accuracy across brain regions revealed that auditory and language-processing areas yielded the highest correlations. In particular, superior temporal regions and the broader temporal cortex demonstrated the strongest predictive performance, highlighting the model’s ability to capture modality-specific neural dynamics.
The transition from ID to OOD evaluation resulted in an expected decline in overall performance, reflecting the substantial shift in stimulus characteristics. However, the spatial pattern of predictable regions remained relatively consistent, with language and auditory areas maintaining higher accuracy compared to visual regions (Figure \ref{fig:combined}). This suggests that our feature extraction approach successfully captured generalizable representations for audio-linguistic processing, while visual features may benefit from a broader range of characteristics represented.
The results indicate that higher-order processing areas involved in complex cognitive functions proved more challenging to predict than primary sensory regions, highlighting the inherent difficulty in modeling abstract neural computations across different stimulus distributions.

\begin{table}[ht]
\centering
\footnotesize
\setlength{\tabcolsep}{1.5pt}
\caption{\textit{Algonauts Project 2025 Challenge leaderboard.} Challenge Score: Pearson correlation between predicted and withheld fMRI responses (a) ID (averaged across parcels, subjects), (b) OOD (averaged across parcels, movies, subjects).}
\label{tab:leaderboards}
\begin{minipage}{0.33\textwidth}
    \centering
    \begin{tabular}{lcc}
    \toprule
    \textbf{Rank} & \textbf{Team} & \textbf{Score} \\
    \midrule
    1  & NCG            & 0.320\\
    2  & sdascoli       & 0.319\\
    3  & SDA            & 0.313 \\
    4  & angelneer926   & 0.296 \\
    5  & CVIU-UARK      & 0.296 \\
    6  & VIL            & 0.295 \\
    7  & MedARC         & 0.288 \\
    8  & ckadirt        & 0.273 \\
    \textbf{9}  & \textbf{corsi01} & \textbf{0.266} \\
    10 & ICL\_SNU       & 0.263\\
    \vdots & \vdots & \vdots \\
    34 & Baseline & 0.203 \\ 
    \bottomrule
    \end{tabular} \vspace{0.1cm}
    \\(a)
\end{minipage}%
\hspace{-1cm}
\begin{minipage}{0.33\textwidth}
    \centering
    \begin{tabular}{lcc}
    \toprule
    \textbf{Rank} & \textbf{Team} & \textbf{Score} \\
    \midrule
    1 & sdascoli       & 0.215 \\
    2 & NCG            & 0.210 \\
    3 & SDA            & 0.209 \\
    4 & ckadirt        & 0.209 \\
    5 & CVIU-UARK      & 0.205 \\
    6 & angelneer926   & 0.199 \\
    7 & ICL\_SNU        & 0.161 \\
    \textbf{8} & \textbf{corsi01} & \textbf{0.158} \\
    9 & alit           & 0.157 \\
    10 & robertscholz   & 0.150 \\
    \vdots & \vdots & \vdots \\
    21 & Baseline & 0.090 \\
    \bottomrule
    \end{tabular}\vspace{0.1cm}
    \\(b)
\end{minipage}
\end{table}

\section{Conclusion}
This work presented a novel approach for predicting brain responses to multimodal stimuli by leveraging the brain’s functional organization. Our network-clustered architecture, based on Yeo's 7-network parcellation, enabled specialized modeling approaches for distinct brain systems and 
incorporated adaptive memory components where beneficial. Key contributions include the development of custom multimodal feature extraction pipelines and the effective scaling of encoding models to whole-brain prediction across 1000 cortical regions. The multi-subject approach successfully captured both shared neural principles and individual differences, achieving competitive performance in the Algonauts 2025 Challenge.
Future directions include exploring more advanced memory mechanisms, expanding the diversity and depth of visual feature representations, and extending the network-clustered modeling approach to other neuroimaging datasets and stimulus modalities.


%



\section*{Acknowledgment}
Financial support from ICSC – Centro Nazionale di Ricerca in High Performance Computing, Big Data and Quantum Computing, funded by European Union – NextGenerationEU.
This work was partially funded by the National Plan for NRRP Complementary Investments (PNC, established with the decree-law 6 May 2021, n. 59, converted by law n. 101 of 2021) in the call for the funding of research initiatives for technologies and innovative trajectories in the health and care sectors (Directorial Decree n. 931 of 06-06-2022) - project n. PNC0000003 - AdvaNced Technologies for Human-centrEd Medicine (project acronym: ANTHEM)\footnote{\url{https://fondazioneanthem.it/}}. This work reflects only the authors’ views and opinions, neither the Ministry for University and Research nor the European Commission can be considered responsible for them.

\end{document}